\documentclass[aps,pra,twocolumn]{revtex4}

\usepackage[usenames,dvipsnames]{color}

\usepackage[latin1]{inputenc}
\usepackage{graphicx}
\usepackage{amssymb}
\usepackage{subfigure}
\usepackage{stackrel}

\usepackage{amsmath}
\usepackage{amsfonts}
\usepackage{bm}
\usepackage{color}
\usepackage{verbatim} 

\newcommand{\ket}[1]{\ensuremath{\left|{#1}\right\rangle}}

\newcommand{\beq}{\begin{equation}}
\newcommand{\eeq}{\end{equation}}
\newcommand{\bse}{\begin{subequations}}
\newcommand{\ese}{\end{subequations}}
\newcommand{\bea}{\begin{eqnarray}}
\newcommand{\eea}{\end{eqnarray}}
\newcommand{\bit}{\begin{itemize}}
\newcommand{\eit}{\end{itemize}}
\newcommand{\bpmatrix}{\begin{pmatrix}}
\newcommand{\epmatrix}{\end{pmatrix}}

\newcommand{\be}{\begin{equation}}
\newcommand{\ee}{\end{equation}}
\newcommand{\ben}{\begin{eqnarray}}
\newcommand{\een}{\end{eqnarray}}

\begin{document}

\title{Problem of quantifying quantum correlations with non-commutative discord}

\author{A. P. Majtey$^{1,2}$, D. Bussandri$^{1,2}$ , T. M. Osán$^{1,3}$, P. W. Lamberti$^{1,2}$, A. Vald\'es-Hern\'andez$^4$}
\affiliation{$^1$Facultad de Matem\'atica, Astronom\'{\i}a, F\'{\i}sica y Computaci\'on, Universidad Nacional de C\'ordoba, Av. Medina Allende s/n, Ciudad Universitaria, X5000HUA C\'ordoba, Argentina}
\affiliation{$^2$Consejo Nacional de Investigaciones Cient\'{i}ficas y T\'ecnicas de la Rep\'ublica Argentina, Av. Rivadavia 1917, C1033AAJ, CABA, Argentina}
\affiliation{$^3$ Instituto de Física Enrique Gaviola, Consejo Nacional de Investigaciones Cient\'{i}ficas y T\'ecnicas de la Rep\'ublica Argentina, Av. Medina Allende s/n, X5000HUA, Córdoba, Argentina}
\affiliation{$^4$Instituto de F\'{\i}sica, Universidad Nacional Aut\'{o}noma de M\'{e}xico, Apartado Postal 20-364, M\'{e}xico D.F., Mexico}


\email[]{amajtey@famaf.unc.edu.ar}

\begin{abstract}
In this work we analyze a non-commutativity measure of quantum correlations recently proposed by Y. Guo [Sci. Rep. {\bf 6}, 25241 (2016)]. By resorting to a systematic survey of a two-qubit system, we detected an undesirable behavior of such a measure related to its representation-dependence. In the case of pure states, this dependence manifests as a non-satisfactory entanglement measure whenever a representation other than the Schmidt's is used. In order to avoid this basis-dependence feature, we argue that a minimization procedure over the set of all possible representations of the quantum state is required. In the case of pure states, this minimization can be analytically performed and the optimal basis turns out to be that of Schmidt's. In addition, the resulting measure inherits the main properties of Guo's measure and, unlike the latter, it reduces to a legitimate entanglement measure in the case of pure states. Some examples involving general mixed states are also analyzed considering such an optimization. The results show that, in most cases of interest, the use of Guo's measure can result in an overestimation of quantum correlations. However, since Guo's measure has the advantage of being easily computable, it might be used as a qualitative estimator of the presence of quantum correlations.
\end{abstract} 

\maketitle

\section{\label{sec:intro}Introduction}
Quantum Information Theory (QIT) is concerned with the use of quantum resources to perform tasks of information processing which are either not feasible to be implemented classically or can be performed with classical devices in a way much less efficient. The fact that in most cases quantum protocols can outperform their classical counterparts (if such a thing is feasible to be done) is generally attributed to the existence of quantum correlations (QCs) \cite{Modi2012,Adesso2016,Szan2015,Sun17,Sun17b,Wang17}. For a long time, QCs  were associated with the existence of entanglement in composite quantum systems. Besides, according to Schr\"odinger himself, entanglement is ``the characteristic trait of quantum mechanics'' \cite{Schrod35a,Schrod35b} and has been extensively studied in connection with Bell's inequalities \cite{Bell}. On one hand, entangled states violating Bell's inequalities~\cite{Bell} contain `non local' features which were initially considered as the necessary quantum resource to achieve a computational speedup over the best classical algorithm~\cite{Nielsen&Chuang}. On the other hand, since (mixed) separable states do not violate Bell's inequalities and can be prepared by local operations and classical communication (LOCC), until very recently they were considered as purely classical and, in consequence, useless for tasks of quantum information processing.
However, further research has provided a great amount of evidence supporting the idea that this is not the case \cite{Knill98,Braun99,Meyer00,Datta05,Datta07,Datta08,Lanyon08}. As a consequence, the study of entanglement measures was extended in order to include the quantification of more general quantum correlations \cite{Modi2012,Adesso2016}.
One of the most widely used measure of quantum correlations in bipartite systems is the 
so-called (standard) \emph{quantum discord} (QD)~\cite{OZ02,HV01}. In few words, QD quantifies the discrepancy between the quantum versions of two classically equivalent expressions for
mutual information. Even though, from a conceptual point of view, QD is of relevance in assessing possible non-classical resources for information processing, for a practical use it presents some drawbacks. For example, at this moment, there is no straightforward criterion to verify the presence of discord in a given general bipartite quantum state  (i.e., a bipartite state belonging to the product of two Hilbert spaces of arbitrary dimensions). As the evaluation of QD involves an optimization procedure, analytical results are known only in some particular cases~\cite{Lang10,Cen11,Adesso10,Ali10,Shi11,Chen11,Lu11,Giro12,Li11,Luo08b}. Furthermore, in general, calculation of quantum discord is NP-complete since the optimization procedure needs to be done sweeping a complete set of measurements over one of the subsystems \cite{Huang14}. 

With the aim of finding a measure of QCs easier to calculate, several alternative measures to QD have been proposed \cite{DVB10,Brod10,Paula13,Spehner14a,Spehner14b,Jakob14}. For example, we can mention discord-like quantities \cite{Brod10}, geometric measures to quantify QD \cite{DVB10,LuoGD10}, and a measure based on Bures distance \cite{Spehner14a,Spehner14b}, among others. In the particular case of qubit-qudit states it is worth mentioning that an interesting discord type measure based on the quantum uncertainty of local single observables, which can be (closed) analytically computed, was introduced in \cite{GTA13}. However, in general, most of the alternative measures of QCs (if not ill-defined) become difficult to calculate since they also involve an optimization process either in a minimization or in a maximization scenario. Additionally, in some cases, undesired behaviors of the measures reduce the potential of their applicability. Furthermore, the great number of measures currently found in literature make it difficult to progress in the study of their properties in order to assure they provide trustful measures of QCs. 
Thus, at present, there is not a general agreement about which measure of QCs is the most suitable to be used in a practical way in an arbitrary composite quantum system. Hence, extreme caution should be exercised in devising new practical methods to quantify QCs in order to avoid undesired behaviors and subtleties in their properties. In summary, it seems that an examination of the properties of any new promising measure of QCs should be carefully addressed. Following this last direction, in this work we investigate various features of a non-commutativity measure of QCs introduced by Guo in a recent work \cite{Guo16}. In that work, Guo introduced two QCs measures in terms of the non-commutativity of some operators quantified by the trace norm and the Hilbert-Schmidt norm. According to \cite{Guo16}, the non-commutative quantum discord (NCQD) measures can be computed directly for any arbitrary state without requiring any previous optimization procedure, as is the case with usual discord. In this work we show that, indeed, the NCQD measures have the drawback of depending upon the representation of the state, and suggest a new measure to overcome this undesirable feature.\\
This paper is organized as follows. In section \ref{sec:NCQD}, we review the definition and main properties of the NCQD measures. In section \ref{sec:repdepPS}, we discuss its representation-dependence feature resorting to computational and Schmidt representations of pure states. We also propose a new measure that is representation-independent and extends to the general (mixed, $d$-dimensional) case. In section \ref{sec:repdepMS}, we calculate the new measure for some typical examples, comparing it with NCQD measure introduced by Guo. Finally, some conclusions are drawn in section \ref{sec:concl}.
 
\section{\label{sec:NCQD}Non-commutativity measure of quantum correlations}

Let us consider a bipartite system ($A+B$) in an arbitrary quantum state $\rho$, defined on the Hilbert space $\mathcal{H}=\mathcal{H}_A\otimes\mathcal{H}_B$. If $\{|i_A\rangle\}$ stands for an orthonormal basis of $\mathcal{H}_A$, then $\rho$ can be represented by

\beq
\rho=\sum_{i,j} |i_A\rangle\langle j_A|\otimes B_{ij},\label{rho1}
\eeq

\noindent where $B_{ij}=\mathrm{Tr}_A[(|j_A\rangle\langle i_A|\otimes \mathbb{I}_B)\rho]$ or, equivalently, $
B_{ij}=\langle i_{A}|\rho|j_{A}\rangle.$

With the operators $B_{ij}$ just defined, Guo \cite{Guo16} introduced two non-commutativity measures as follows:

\beq
D_G(\rho):=\sum_{\Omega}||[B_{ij},B_{kl}]||_{\mathrm{Tr}},\label{DG}
\eeq

\noindent and

\beq
D'_G(\rho):=\sum_{\Omega}||[B_{ij},B_{kl}]||_2,\label{DG'}
\eeq

\noindent where $\Omega$ represents the set of all the possible pairs (regardless of the order), and $||\cdot||_{\mathrm{Tr}}$ and $||\cdot||_2$ denote, respectively, the trace and the Hilbert-Schmidt norm, i.e., $||A||_{\mathrm{Tr}}=\mathrm{Tr}(\sqrt{A^{\dagger}A})$ and $||A||_2=\sqrt{\mathrm{Tr}(A^{\dagger}A)}$.

Resorting to the fact that (usual) quantum discord $D(\rho)$ vanishes if and only if all the ($B$)-local operators $B_{ij}$ are mutually commuting normal operators \cite{DVB10,GH12}, Guo proposes the non-commuting measures \eqref{DG} and \eqref{DG'} as measures of quantum discord. As explained in \cite{Guo16},
$D_G(\rho)$ and $D'_G(\rho)$ satisfy the following properties: (i) $D_G(\rho)=D'_G(\rho)=0$ iff $D(\rho)=0$; (ii) $D_G(\rho)$ and $D'_G(\rho)$ are invariant under local unitary operations, i.e., $D_G(\rho)=D_G(U_A\otimes U_B\rho\, U_A^{\dagger}\otimes U_B^{\dagger})$ and $D'_G(\rho)=D'_G(U_A\otimes U_B\rho\, U_A^{\dagger}\otimes U_B^{\dagger})$, being $U_A$ and $U_B$ arbitrary unitary operators in $\mathcal{H}_A$ and $\mathcal{H}_B$, respectively. Unlike usual quantum discord and other measures of non-classicality, according to Guo \cite{Guo16}, the proposed measures are not based on measurements performed on one of the subsystems, and can be computed directly for any arbitrary state without requiring any previous optimization procedure. Note however that the Hilbert-Schmidt norm is easier to calculate (compared with the trace norm), hence from now on we will focus on the Hilbert-Schmidt norm measure, $D'_G(\rho)$.

\section{\label{sec:repdepPS}Pure states. Representation-dependence of the noncommutativity measure}\label{SecPure}

From Eq. (\ref{rho1}) it follows that the operators $B_{ij}$ can be identified with blocks of the matrix $\rho$. In a two-qubit system, for example, \beq
\rho=
\begin{pmatrix} \label{rhoB}
      B_{00} & B_{01} \\
      B_{10} & B_{11} 
   \end{pmatrix},
\eeq

\noindent where 0 and 1 denote each of the two (orthonormal) basis vectors $\{|i_A\rangle\}$, put in correspondence with the canonical basis $\ket{\epsilon_{0}}=(1,0)^{\textit T}$, and $\ket{\epsilon_{1}}=(0,1)^{\textit T}$. Now, let us consider a two-qubit system in an arbitrary pure state

\beq
|\psi\rangle=a|00\rangle+b|01\rangle+c|10\rangle+d|11\rangle,
\eeq

\noindent where $a$, $b$, $c$, and $d$ are complex numbers that satisfy  $|a|^2+|b|^2+ |c|^2 +|d|^2=1$. Thus, the density matrix $\rho$ is given by Eq. (\ref{rhoB}), where

\beq
B_{00}=
\begin{pmatrix} 
      |a|^2 & ab^* \\
      a^*b & |b|^2 
   \end{pmatrix},
\eeq

\beq
B_{01}=
\begin{pmatrix} 
      ac^* & ad^* \\
      bc^* & bd^* 
   \end{pmatrix},
\eeq

\beq
B_{10}=
\begin{pmatrix} 
      a^*c & b^*c \\
      a^*d & b^*d 
   \end{pmatrix},
\eeq

\noindent and

\beq
B_{11}=
\begin{pmatrix} 
      |c|^2 & cd^* \\ 
      c^*d & |d|^2 
   \end{pmatrix}.
\eeq
After a direct calculation of the six commutators $[B_{00}, B_{01}]$, $[B_{00}, B_{10}]$, $[B_{00}, B_{11}]$, $[B_{01}, B_{10}]$, $[B_{01}, B_{11}]$, and $[B_{10}, B_{11}]$, the measure $D'_G(\rho)$ can be written as

\begin{equation}\label{Dcomp}
D'_G(\rho)=C\Big[1+\frac{1}{2\sqrt{2}}\Big(\sqrt{C^2+4|(\rho_A)_{01}|^2}+2 |(\rho_A)_{01}|\Big)\Big],
\end{equation}

\noindent where $\rho_A=\textrm{Tr}_{B}\rho$ is the (reduced) density matrix corresponding to subsystem $A$. Therefore, $|(\rho_A)_{01}|=|a^*c+b^*d|$ represents a measure of the coherence of $\rho_A$, and $C=2 |ad-bc|$ stands for Wootters' concurrence \cite{W98} which is a measure of the entanglement between $A$ and $B$.

The fact that $D'_G(\rho)$ depends upon a parameter related to $\rho_A$ ensues from the fact that $\rho$ has been decomposed in the form (\ref{rho1}), associated with the bipartition $A|B$. The bipartition $B|A$ corresponds to the decomposition [cf. Eq. \eqref{rho1}] 

\beq
\rho=\sum_{i,j} A_{ij}\otimes|i_B\rangle\langle j_B|,\label{rho2}
\eeq

\noindent where $A_{ij}=\mathrm{Tr}_B[(\mathbb{I}_A \otimes |j_B\rangle\langle i_B|) \rho]$ or, equivalently, $A_{ij}=\langle i_{B}|\rho|j_{B}\rangle$. If instead of considering the bipartition $A|B$ we consider the bipartition $B|A$, the term $(\rho_A)_{01}$ in Eq. (\ref{Dcomp}) must be replaced by $(\rho_B)_{01}=|a^*b+c^*d|$, i.e., the coherence of the reduced density matrix $\rho_B$. This means that the correlation between $A$ and $B$ and the correlation between $B$ and $A$, as measured by $D'_G(\rho)$, do not coincide in general. This is an undesirable feature since, even though discord is known to be a non-symmetric measure of quantum correlations, for pure states --as the one considered here-- it should reduce to an entanglement measure which should be symmetric under the exchange $A\leftrightarrow B$ (e.g., Wootters' concurrence $C$). On the other hand, the coherence term present in Eq. (\ref{Dcomp}), unlike the concurrence $C$, depends upon the specific representation of $\rho$. Thus, $D'_G(\rho)$ becomes representation-dependent, which is another undesirable property for an entanglement measure. Therefore, we are led to conclude that for pure states $D'_G(\rho)$ does not reduce to a good entanglement measure (i.e., symmetric and representation-independent). This fact puts at stake its adequacy when dealing with general (mixed) states. In what follows we discuss how these disadvantages can be surmounted.

According to the aforementioned observations, it is precisely the coherence term appearing in Eq. (\ref{Dcomp}) what introduces the inconvenient properties in $D'_G(\rho)$. Thus, as a first step, we require that $(\rho_A)_{01}$ \textit{and} $(\rho_B)_{01}$ shall reduce to zero for \textit{all} $\rho=\rho^2$. This condition is met only when both $\rho_A$ and $\rho_B$ are both diagonal, which holds irrespective of the state whenever $\ket{\psi}$ is decomposed into its Schmidt form

\beq \label{psiS}
|\psi\rangle=\sum_{n=0,1} \sqrt{\lambda_n} |v^{A}_n\rangle\otimes |u^{B}_n\rangle,
\eeq

\noindent where $\lambda_n$ stands for the eigenvalues of $\rho_A$ and $\rho_B$, so that $\lambda_0+\lambda_1=1$, and $\{|v^{A}_n\rangle\}$, $\{|u^{B}_n\rangle\}$ are the corresponding (orthonormal) eigenvectors. Thus, the measure $D'_G$ reduces (in the Schmidt representation) to

\begin{equation}
D'_G(\rho_{\textrm{Sch}})=2\sqrt{\lambda_0 \lambda_1}+\sqrt{2}\lambda_0 \lambda_1,
\end{equation}

\noindent in agreement with Guo's result. Using the fact that $C^2=4\lambda_0 \lambda_1$, we get

\begin{equation}\label{Dsch}
D'_G(\rho_{\textrm{Sch}})=D'_G(\rho)|_{(\rho_A)_{01}=0}=C\left(1+\frac{1}{2\sqrt{2}}C\right),
\end{equation}

\noindent as follows from Eq. (\ref{Dcomp}) with $(\rho_A)_{01}=0$. Thus, resorting to the Schmidt representation (the only one considered in \cite{Guo16}), we see that $D'_G(\rho)$ reduces to a monotonic function of concurrence --hence to a measure of entanglement--, whose maximum value is attained for maximally entangled states (i.e., $C=1$). However, in any other representation this ceases to be the case (of course, this is due to the coherences aforementioned). 
In addition, it is straightforward to see that Eq. (\ref{Dcomp}) (or the analogous equation corresponding to the bipartition $B|A$) is minimized, with $C$ fixed, for $(\rho_A)_{01}=0$ (or $(\rho_B)_{01}=0$). This means that Eq. (\ref{Dcomp}) attains its minimum (and symmetrical) value whenever $\rho$ is expressed in its Schmidt representation, whence   

\begin{equation}
D'_G(\rho_{\textrm{Sch}})=\min D'_G(\rho),
\end{equation} 
where the minimum is taken over all decompositions of $\ket{\psi}$. Consequently, the quantity

\begin{equation}
d'(\rho):=\min_{\cal{R}} D'_G(\rho),\label{d}
\end{equation}

\noindent being $\cal{R}$ the set of all possible representations of $\rho$, constitutes a non-commutativity measure that inherits the main properties of Guo's measure $D'_G$ but, unlike the latter, reduces to a legitimate entanglement measure for pure states. Notice that $d'(\rho)$ applies for general (mixed and pure) states of bipartite systems of arbitrary dimensions. However,  this measure requires a minimization procedure --as is the case with usual discord-- which can be difficult to calculate for general mixed states. In the next section we analyze some examples involving the evaluation of $d'(\rho)$ in the case of mixed states.

\begin{figure}[h]
\begin{center}
\includegraphics[scale=0.32,angle=0]{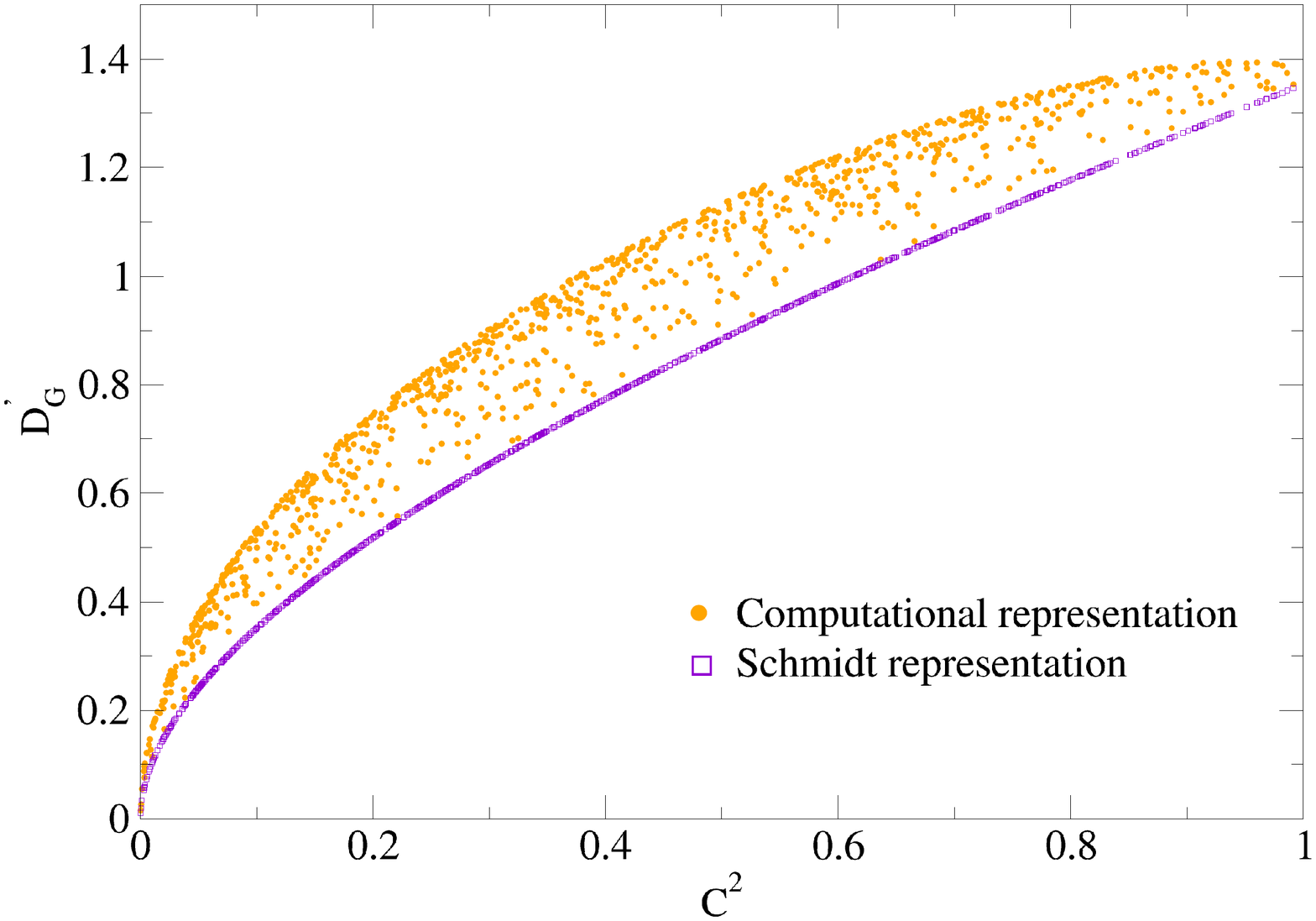}
\caption{(Color online). Representation-dependence of $D'_G(\rho)$ for $10^6$ randomly generated two-qubit pure states as a function of the square of the concurrence $C$. Orange symbols (circles) correspond to $D'_G$ in the computational representation, and purple symbols (squares) to the Schmidt basis. All plotted quantities are dimensionless. \label{purerepre}}
\end{center}
\end{figure}

In Fig. \ref{purerepre} we show the effect of the representation-dependence of  $D'_G$. We generated $10^6$ two-qubit random pure states distributed according to the Haar measure \cite{ZHS98,PZK98} and computed $D'_G(\rho)$ using both, the computational and the Schmidt representations. We plotted $D'_G$ as a function of the square of the concurrence $C$. Notice that, as expected,  the values obtained in the Schmidt representation (purple squares) are in all cases lower than those corresponding (for the same state) to the computational representation (orange dots). The maximum value of $D'_G(\rho_{\textrm{Sch}})$ is 1.3535 and corresponds to states with $C=1$, though $D'_G(\rho_{\textrm{Comp}})$ attains its maximum value (1.3964), i.e., not for a maximally entangled state but for a state with concurrence $C=0.9725$.

Finally, notice that for  states $\ket{\psi}$ maximally entangled, the reduced density matrices $\rho_A$ and $\rho_B$ do coincide and the coherence terms reduce to zero. Hence, for maximally entangled pure states \textit{any} decomposition will display the same value of $D'_G(\rho)$.


\section{\label{sec:repdepMS}Representation-independent measure. Some examples for mixed states.}

In Ref. \cite{Guo16}, $D_G$ and $D'_G$  were computed and compared with the usual discord for several families of mixed states, namely Werner, isotropic, and Bell-diagonal states. Here we will focus our discussion on general (mixed) states of two qubits. In what follows, we briefly discuss how $D'_G(\rho)$ depends upon the representation, and compare it with the measure $d'(\rho)$ introduced earlier in Sec. \ref{SecPure} [cf. Eq. \eqref{d}].

Let $|i'_A\rangle=U_A|i_A\rangle$ be an arbitrary orthonormal basis of $\cal{H}_A$. Then, the state $\rho$ can also be represented as [cf. Eq. \eqref{rho1}]

\begin{equation}
\rho=\sum_{ij} |i'_A\rangle\langle j'_A|\otimes B'_{ij}.
\end{equation}
In this new representation, the operators $B'_{ij}$ take the form

\begin{eqnarray}
B'_{ij}&=&\textrm{Tr}_A\{(|j'_A\rangle\langle i'_A|\otimes \mathbb{I}_B)\rho\}\nonumber\\
&=&\langle i'_A|\rho|j'_A\rangle=\langle i_A|U_A^{\dagger}\rho\, U_A |j_A\rangle,
\end{eqnarray}
hence, the $B'_{ij}$ can be identified with the block components of the matrix [cf. Eq. \eqref{rhoB}]

\begin{eqnarray}
U_A^{\dagger}\rho U_A=\begin{pmatrix} \label{rhoBprime}
      B'_{00} & B'_{01} \\
      B'_{10} & B'_{11} 
   \end{pmatrix}.
\end{eqnarray}

Now, with these primed operators we can compute $D'_G$ in the new representation. The optimization procedure in (\ref{d}) is then reduced to search the minimum of $D'_G$ over the set of all possible basis of $\cal{H}_A$, that is, over all transformations belonging to $SU(2)$ \cite{BZ06}.

\begin{figure}[h]
\begin{center}
\includegraphics[scale=0.32,angle=0]{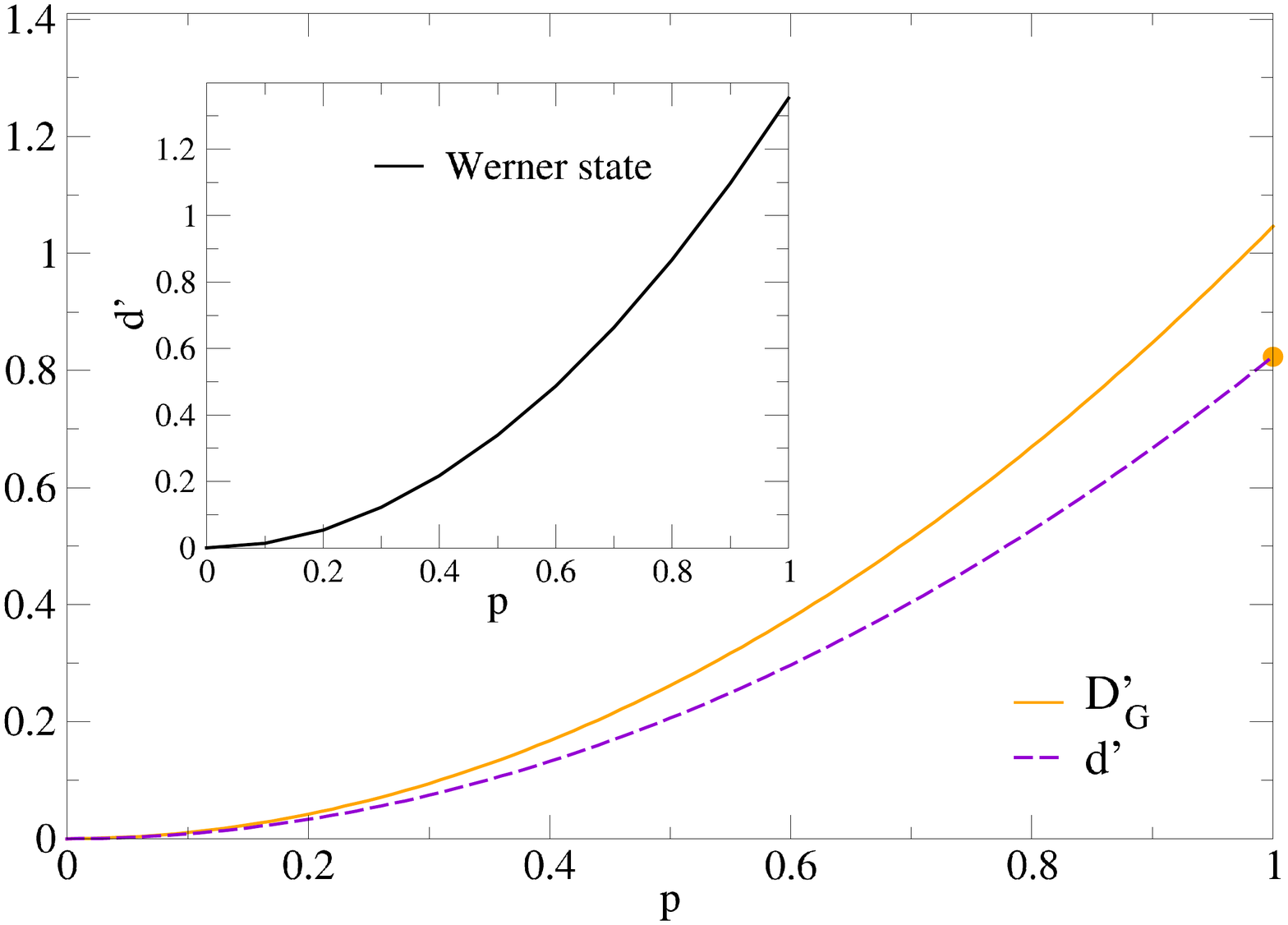}
\caption{(Color online). $D'_G(\rho)$ in the computational basis (solid orange line), and $d'$ (dashed purple line) as a function of the parameter $p$ for the state (\ref{ex1}), with $|\psi\rangle=1/\sqrt{3}(|00\rangle+|01\rangle+|10\rangle)$. The orange dot represents the measure for the pure state in the Schmidt representation. The minimization was performed over arbitrary representations of $\rho$ associated to orthonormal local basis. Inset: $D'_G(\rho)$ for a Werner state, i.e., the state given by Eq.\eqref{ex1} with $|\psi\rangle$ the Bell state $\beta_{00}$ (see below Eq. \eqref{rhobell2}). In this case the measure is representation-independent. All plotted quantities are dimensionless.\label{pstate}}
\end{center}
\end{figure}

In order to be more general than in \cite{Guo16} we will consider a mixed state $\rho$ of the form
\begin{equation}\label{ex1}
\rho=(1-p)\frac{\mathbb{I}}{4} +p|\psi\rangle\langle \psi|,
\end{equation}

\noindent where $|\psi\rangle$ is any arbitrary pure state. If $|\psi\rangle$ corresponds to a Bell state, the state $\rho$ becomes symmetric under the interchange of the subsystems, the measure $D'_G$ becomes symmetric in both bipartitions ($A|B$ and $B|A$), and also representation-independent (see inset of Fig. \ref{pstate}). However, if we take for instance $|\psi\rangle=1/\sqrt{3}(|00\rangle+|01\rangle+|10\rangle)$, the measure becomes dependent upon the representation. In Fig. \ref{pstate} we show this fact explicitly by plotting $D'_G$ as a function of $p$ considering the computational representation (solid curve), and the measure $d'$ introduced in Eq. \eqref{d} (dashed curve). As $p$ goes from $p=0$ to $p=1$ (i.e., as the state (\ref{ex1}) goes from being a maximally mixed state to a pure state) the difference between $D'_G$ and $d'$ increases continuously. At $p=1$, $d'$ reduces to $D'_G(\rho_{\textrm{Sch}})$ (orange dot), i.e., to the measure considered in Ref. \cite{Guo16} for pure states. Nevertheless, $D'_G$ does not coincide with such a value, which means that, unless the minimization in Eq. (\ref{d}) is performed, the measure $D'_G(\rho)$ will in general exhibit discontinuities when a mixed state transforms into a pure one. 

%

\begin{figure*}[htb]
\begin{minipage}[b]{.4\linewidth}
\begin{center}
 \subfigure[(Color online). $D'_G$ (solid orange line) and $d'$ (dotted blue line) as a function of
$p$ for $\rho_1$. All plotted quantities are dimensionless.]{
   \includegraphics[scale =0.2] {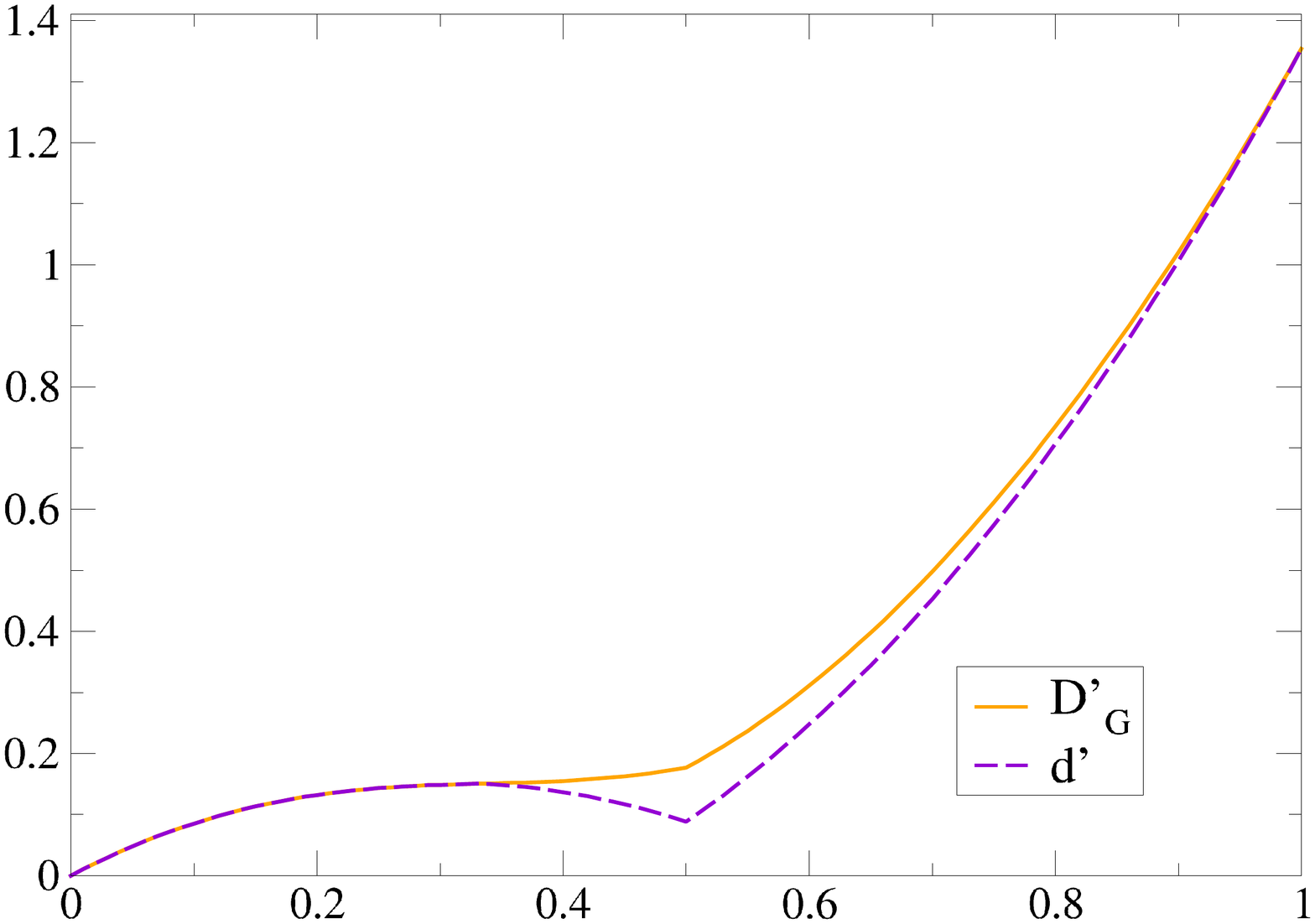}
   \label{subfig1}
 }
\end{center}
\end{minipage}\hfill
\begin{minipage}[b]{.5\linewidth}
\begin{center}
\vspace{-0.2cm}
 \subfigure[(Color online). $D'_G$ (solid orange line) and $d'$ (dotted purple line) as a function of
$p$ for $\rho_2$. All plotted quantities are dimensionless.]{
   \includegraphics[scale =0.2] {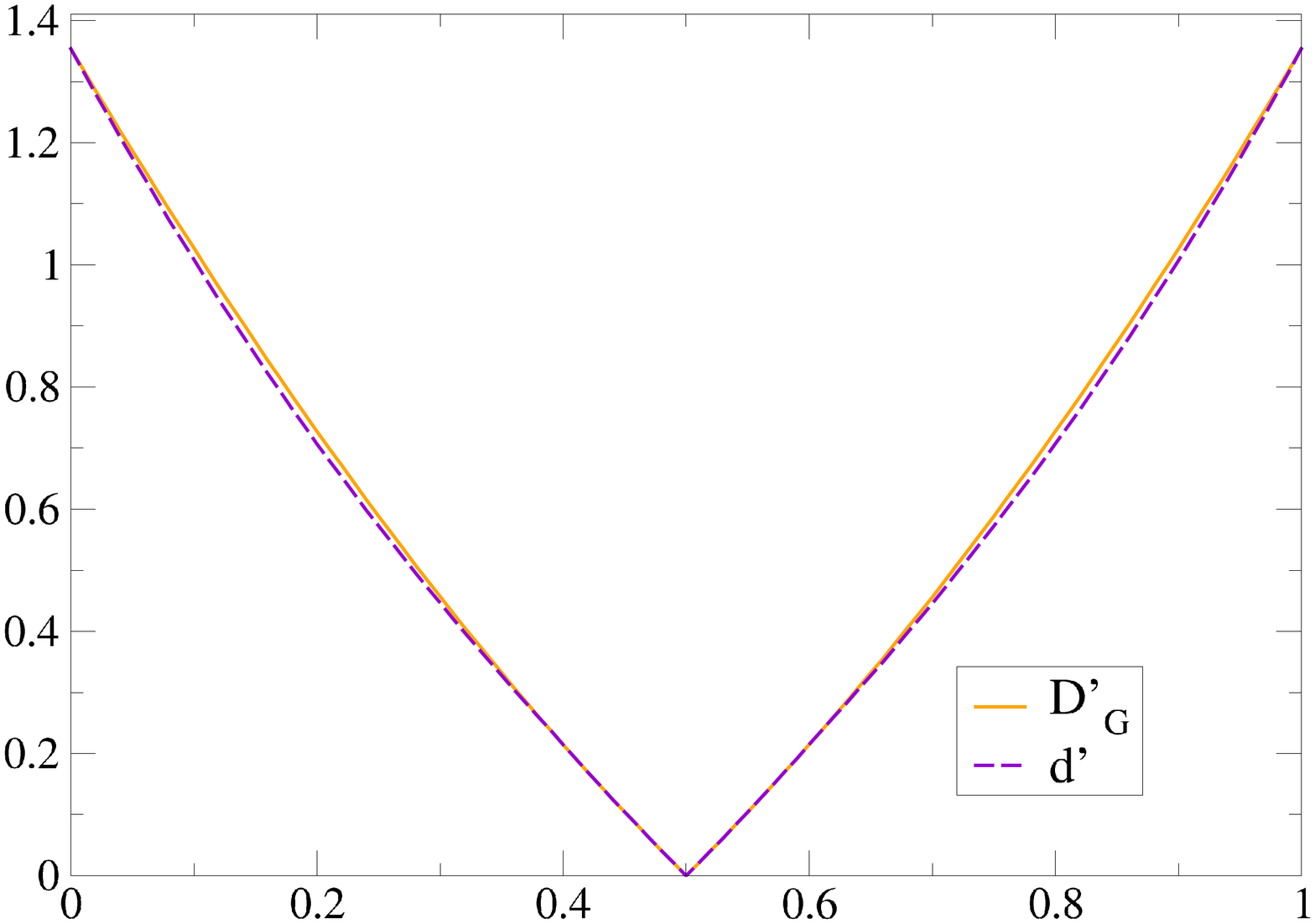}
   \label{subfig2}
 }
\end{center}
\end{minipage}
\caption{The graphs pertain to the states given by
Eqs.(\ref{rhobell1}) and (\ref{rhobell2}). }
\end{figure*}

As a second example we will compute $d'$ for some of the Bell-diagonal states analized in Ref. \cite{Guo16}. Specifically we will consider the states

\begin{equation}\label{rhobell1}
\rho_1=p|\beta_{11}\rangle\langle\beta_{11}|+\frac{1-p}{2}(|\beta_{01}\rangle\langle\beta_{01}|+|\beta_{00}\rangle\langle\beta_{00}|),
\end{equation}
and 
\begin{equation}\label{rhobell2}
\rho_2=p|\beta_{11}\rangle\langle\beta_{11}|+(1-p)|\beta_{01}\rangle\langle\beta_{01}|,
\end{equation}
where $\{|\beta_{ab}\rangle\}$ are four Bell states $|\beta_{ab}\rangle\equiv\frac{1}{\sqrt 2}[|0,b\rangle+(-1)^a|1, 1\oplus b\rangle]$.

In Figs. \ref{subfig1} and \ref{subfig2} we plotted $D'_G$ in the computational representation as in Ref. \cite{Guo16}, and $d'$ as a function of the parameter $p$ for the states $\rho_1$ and $\rho_2$, respectively. For both states the optimization was numerically performed on $SU(2)$. Here, unlike the previous example, in both cases $D'_G$ and $d'$ do coincide for $p=1$, i.e., when the states become pure states. Of course, this is so because for $p=1$ both $\rho_1$ and $\rho_2$ reduce to a maximally entangled pure states for which all representations give the same measure (see last paragraph in Section \ref{SecPure}). However, for intermediate values of $p$, $D'_G(\rho_1)$ and $D'_G(\rho_2)$ overestimate the amount of QCs. For $\rho_1$ the difference is quite significant and the larger discrepancy is reached at $p=0.5$.  Thus, $D'_G(\rho_2)$ works as a tight upper bound for $d'(\rho_2)$ and at $p=0.5$ both measures do coincide. This is so because for $p=0.5$  $\rho_2$ is classically correlated (or zero discordant). Note that $\rho_2$ becomes a pure state also when $p=0$. Although $D'_G$ nearly coincides with $d'$ for $\rho_2$, our examples show that the measure is still representation-dependent when arbitrary mixed states are considered.
At this point, it is important to realize that, given a state $\rho$ represented in \textit{a given fixed basis}, the measure $D'(\rho)$ will be invariant under local unitary operations. However, given two different representations of the state $\rho$, the two values of $D'(\rho)$ will be in general different from each other.

\section{\label{sec:concl}Concluding remarks}

In this work, we analyzed the measure  $D'_G$ of quantum correlations recently proposed by Y. Guo in Ref. \cite{Guo16}. The measure  $D'_G$ is based on the amount of non-commutativity quantified by the Hilbert-Schmidt norm. Our results show that, in general, the measure $D'_G$ depends upon the representation of the state. First, we focused our study on pure states and, by resorting to the computational representation, we showed that $D'_G$ is a function of both, Wootters' concurrence $C$ of the pure state and the coherence of the reduced density matrix. In addition, due to this latter dependence, $D'_G$ becomes a representation-dependent quantity which, in most cases of interest,  yields different results when the bipartition $A|B$ or $B|A$ is considered. These are undesirable features for any measure of QCs in pure states, since the measure does not reduce to a good measure of entanglement. Based on this findings, in order to overcome this undesirable behavior, we suggested an alternative measure $d'$, which inherits the main properties of Guo's measure $D'_G$.  The proposed measure $d'$ involves a minimization procedure over the set of all local basis that, in the case of pure states, can be analytically performed. In that case, the optimal representation turns out to be that of Schmidt. In addition, unlike $D'_G$, $d'$ reduces to a legitimate entanglement measure in the case of pure states. Next, we numerically computed the new measure $d'$ for some typical arbitrary (mixed) states and explicitly showed that also for mixed states $D'_G$ is representation-dependent. As a consequence, in most cases of interest, our results indicate that the use of $D'_G$ can result in an overestimation of quantum correlations. Nevertheless, regarding arbitrary mixed states, it is worth to mention that the optimization procedure involved in the calculation of $d'$ can be difficult to perform. As a final comment, we would like to point out that, since the NCQC measure introduced by Guo has the advantage of being easily computable, it might be used as a qualitative estimator of the presence of quantum correlations.

\begin{acknowledgements}
A.P.M., D.B., T.M.O, and P.W.L. acknowledge the Argentinian agency SeCyT-UNC and CONICET for financial support. D. B. has a fellowship from CONICET. A.V.H. gratefully acknowledges financial support from DGAPA, UNAM through project PAPIIT IA101816.
\end{acknowledgements}


{}

%
%

\end{document}